\documentclass[%
 reprint,
 amsmath,amssymb,
 aps,
pra]{revtex4-1}

\usepackage{graphicx}
\usepackage{hyperref}

\DeclareMathOperator*{\argmax}{arg\,max}
\DeclareMathOperator*{\argmin}{arg\,min}
\begin{document}         



\title{X-ray optics and beam characterisation using random modulation: Theory}


\author{Sebastien Berujon}
\email{xebastien@berujon.org}
\affiliation{European Synchrotron Radiation Facility, CS 40220, F-38043 Grenoble cedex 9, France}
\author{Ruxandra Cojocaru}
\affiliation{European Synchrotron Radiation Facility, CS 40220, F-38043 Grenoble cedex 9, France}
\author{Pierre Piault}
\affiliation{European Synchrotron Radiation Facility, CS 40220, F-38043 Grenoble cedex 9, France}
\author{Rafael~Celestre}
\affiliation{European Synchrotron Radiation Facility, CS 40220, F-38043 Grenoble cedex 9, France}
\author{Thomas Roth}
\affiliation{European Synchrotron Radiation Facility, CS 40220, F-38043 Grenoble cedex 9, France}
\author{Raymond Barrett}
\affiliation{European Synchrotron Radiation Facility, CS 40220, F-38043 Grenoble cedex 9, France}
\author{Eric Ziegler}
\affiliation{European Synchrotron Radiation Facility, CS 40220, F-38043 Grenoble cedex 9, France}




\date{\today}

\begin{abstract}
X-ray near-field speckle-based phase-sensing approaches provide efficient means to characterise optical elements. Here, we present a theoretical review of several of these speckle methods in the frame of optical characterisation and provide a generalization of the concept. As we also demonstrate experimentally in another paper \cite{Berujon2019b}, the methods theoretically developed here can be applied with different beams and optics and within a variety of situations where at-wavelength metrology is desired. By understanding the differences between the various processing methods, it is possible to find and implement the best suited approach for each metrology scenario.  
\end{abstract}

\maketitle

\section{Introduction}

With wavelengths on the order of the \r{A}ngstr\"{o}m or shorter, hard X-rays put stringent requirements on the optics needed to tailor and transport the beams produced by advanced accelerator-based light sources such as synchrotrons or X-ray free-electron lasers (XFEL). In order to preserve the intrinsic properties of such radiation - such as the temporal and spatial coherence, the time stability, the flux and spectral characteristics - specific and difficult-to-manufacture optics must be employed. The material and the shape of X-ray optics must be controlled at a scale approaching the working wavelength in order to fulfill the Rayleigh criterion and not degrade the beam properties \cite{attwood2017}. To achieve diffraction limited focusing performance, state-of-the-art hard X-ray mirrors can now be manufactured with reflective surfaces offering height errors less than a nanometer from the theoretical perfect elliptical figure \cite{mimura2009,yabashi2014}. Refractive X-ray optics working in transmission have shape error requirements approximately three orders of magnitude less demanding than for reflective optics. In such optics the physical curvature has to be much larger (implying much smaller radii for lens surfaces) to obtain equivalent focusing properties, making the manufacture of large numerical aperture optics equally challenging \cite{roth2017}.

Optical metrology instruments such as long trace profilers, nano-optical machines and interferometers are invaluable tools assisting in the design and fabrication of ever better performing reflective X-ray optics. Nevertheless, at-wavelength metrology of X-ray optics is now considered to be the ultimate tool for characterising the performance of optics at synchrotron and XFEL beamlines. The optimum operating parameters can be attained only by measuring the state of the X-ray beam under working conditions. In addition to providing the necessary input for the alignment of optical components, online metrology can also yield information regarding the beam vibrational stability, the thermal load and the mechanical constraints on the optics.

For the most demanding applications, at-wavelength metrology methods should be able to reach a sensitivity of the order of a hundredth of a wavelength or better with micrometre scale spatial resolution. This corresponds to an angular sensitivity for X-rays better than tens of nanoradians combined with micrometric resolution. While such a metrology performance makes its implementation difficult, the weak interaction of X-rays with matter renders the task even more challenging.

Current state-of-the-art grazing incidence X-ray mirrors for use with coherent illumination must typically present a figure accuracy better than one nanometer over several tens of millimeters \cite{soufli2008}. Meter long mirrors in operation at the European XFEL now fulfill these characteristics ~\cite{vannoni2016}. In contrast, when compound refractive lenses (CRL) working in transmission are stacked together, the accumulated shape error must remain below a few micrometers \cite{seiboth2017}. For such transmission optics the individual characterisation of each of the lenses can provide precious information not only for the shape errors but also for the sub-surface material structure (inclusions, porosity etc). Thus, in view of manufacturing X-ray optics with better performance, online metrology is expected to provide the ultimate information beyond the limitations of offline metrology. Apart from the advantages of measuring the optics under working conditions, the much shorter wavelength of the X-rays compared to visible light provides greater sensitivity and resolution~\cite{sawhney2013}.

Active optics is another application where online metrology is particularly powerful. Many X-ray optics can be optimized using different methods for wavefront correction including the insertion of a phase plate or of a deformable optics. Within the X-ray reflective optics domain this is done using dynamically deformable mirrors. These optical component surfaces can be distorted, up to a certain extent, to tune and shape a beam. This optimization can be achieved more readily if reliable at-wavelength metrology methods are made available~ \cite{satoshi2016}.

A variety of online metrology approaches were developed at synchrotrons over the years. As the quality of the X-ray optics kept increasing, the need for a better metrology was felt more strongly. Although the pencil beam approach \cite{hignette1997} probably remains today the most widespread technique due to its simplicity of implementation, limitations arise from its limited spatial resolution. Additionally, with pulsed sources such as the new X-ray free electron lasers and as well as very low emittance storage rings, online metrology performed out of focus is highly preferred since the flux density at the focal point can easily damage even the most robust detector.

Many other approaches exist and are sometimes implemented for more complete characterisations \cite{weitkamp2005,mercere2005,engelhardt2005,brady2006,revesz2007,kewish2010,wang2011,matsuyama2012,sutter2012,kayser2017,liu2018}. However, at-wavelength metrology has only recently been implemented at synchrotrons as a systematic and routine inspection tool for optical components.

X-ray near-field speckle-based phase-sensing methods are amongst the most recent approaches introduced for online X-ray at-wavelength metrology. Some important advantages of these methods have already been demonstrated, such as the high angular sensitivity combined with an attractive simplicity of experimental implementation. Using partially coherent light from a synchrotron source, these methods have been shown to be fast, efficient and accurate \cite{berujon2016} while the feasibility of applying the methods at laboratory sources has also been validated \cite{zanette2014}.

Here, we review the fundamental differences between the available processing schemes operating in real space for at-wavelength metrology.


\section{Theory}

This section reviews the fundamental principles of X-ray speckle-based metrology and the various numerical processing approaches developed over the years. The recovery of the scattering signal, also accessible with these speckle-based techniques, will not be discussed here, since it is of reduced interest for metrology applications. Conversely, the recovery of the wavefront phase signal allows for the determination of proper focusing and wavefield propagation of the X-rays.

\subsection{Metrology modes}

We define two different two kinds of metrology, or to be more precise, two types of characterisation, described further as the "absolute" and "differential" modes.

\subsubsection{Differential metrology\label{sec:diffmetro}}
Differential metrology is comparable to what is sensed when performing phase contrast imaging. In this method only the phase shift induced by a sample or optical element is measured. For this approach, two data sets are collected, one with the sample present in the beam and one without; the wavefront difference is inferred from these two data sets. Each data set is either a single detector image, or a series of detector images that have been taken at different positions of the speckle generator, moved perpendicularly to the beam.

\begin{figure}
\includegraphics{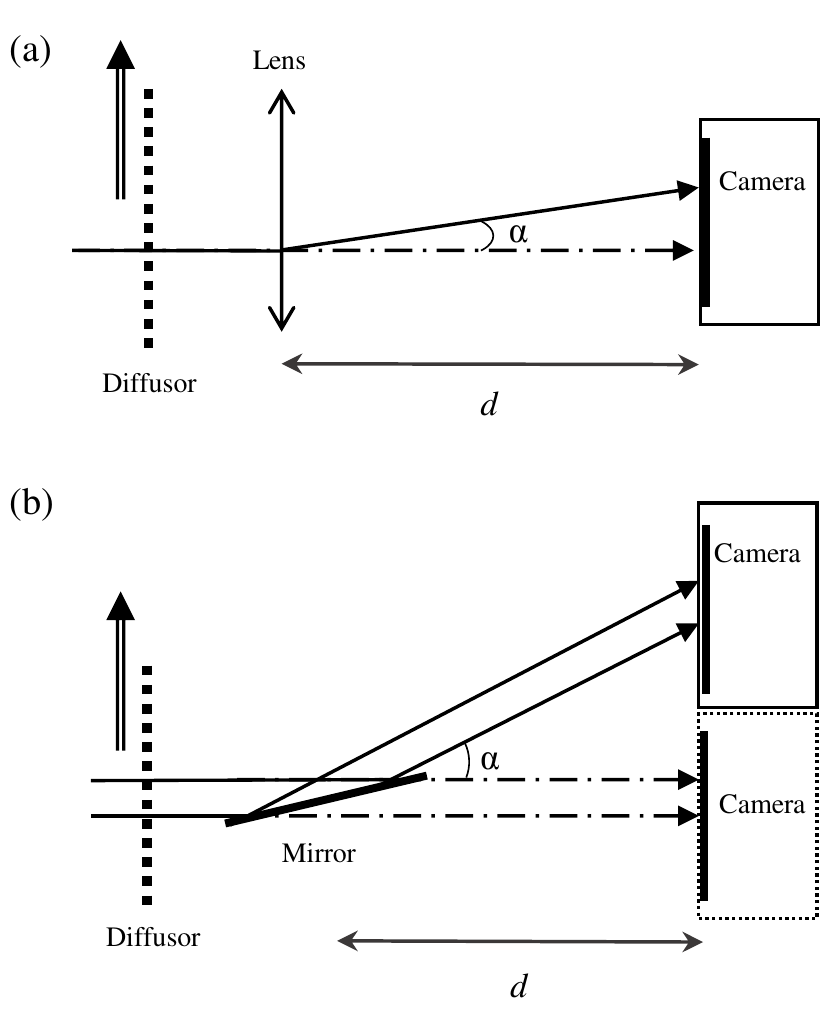}
\caption{ (a) Measurement configuration for the optical characterisation of a lens in transmission. The detector is fixed at all times. (b) Configuration for the optical characterisation of a mirror upon reflection. Here, the detector can be moved transversely to the beam propagation direction between scans, in order to intercept the reflected X-ray beam.\label{fig:geomdiff}}
\end{figure}

Figure~\ref{fig:geomdiff} shows two configurations used for the recovery of the wavefront in differential mode for an object (a)~in transmission and (b)~in reflection. For such measurements, the speckle generator, an object with small phase-shifting features, is generally located upstream of the sample, i.e. the optics under investigation. The propagation distance $d$ used for the calculation of the deflection angle generated by a wavefront distortion is the distance from the sample to the detector.
Alternatively, the scattering screen can be placed downstream of the sample provided that the magnification factor is correctly taken into account in the calculation (for focusing optics).

\subsubsection{Absolute metrology}

Absolute metrology is sensitive to the full state of a beam, that includes the contributions of the source and of all optical elements located in the beam path. In other words, absolute metrology permits the measurement of the X-ray beam's wavefront at any position, for instance at a sample position or downstream of a set of optical elements. Such metrology is necessary to optimize the optical configuration of a beamline and also to provide reference values when comparing different beams. To perform the optimization of active optics, absolute metrology is required to measure the wavefront and eventually correct it.

Typical information obtained using absolute metrology are the radius of curvature of the beam and the optical aberrations. Such data permit the measurement of the focal length and of the beam size at the focal plane.

\begin{figure}
\includegraphics{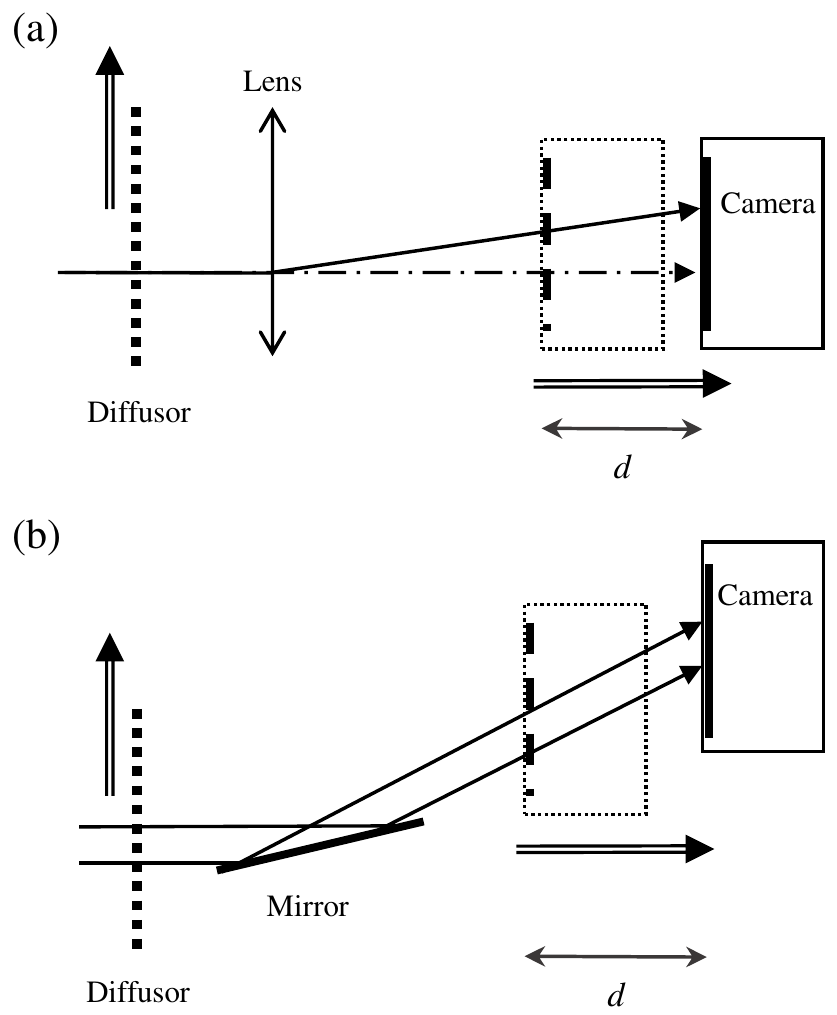}
\caption{Data collection for the absolute mode (a) in-line and (b) upon reflection of the X-rays on an optical element. In both cases the detector collects images in two different propagation planes separated by a typical distance $d$ of about 10~mm to 1000~mm. \label{fig:geomabs}
}
\end{figure}

Figure~\ref{fig:geomabs} shows different geometries employed for absolute metrology. For the X-ray speckle tracking (XST, see Sec.~\ref{sec:XST}), X-ray speckle vector tracking (XSVT, see \ref{sec:xsvt}) and X-ray speckle scanning (XSS, presented in Sec.~\ref{sec:XSS}) methods in absolute mode, the detector has to be moved along the beam axis between two data set acquisitions, $d$ being the distance between the two detector planes. However, for the XSS in the self-correlation mode (described in Sec.~\ref{sec:XSSself}) where a single scan is needed, and similarly for the differential mode, the distance $d$ used for the calculation is the one from the sample to the detector position noted $d$ in Fig.~\ref{fig:geomdiff}.

\subsection{Principles}

\subsubsection{Generalities}

The techniques explained in this manuscript are expected to be applicable over a wide spectral range, but they become particularly advantageous in the X-ray regime as the shorter wavelength leads to an extended near-field range. Indeed, near-field speckle was demonstrated to not change in size and shape over a propagation distance $D\xi/\lambda$, $\lambda$ being the wavelength, $\xi$ the size of the random phase modulator (e.g. the 'grain' size) and $D$ the transverse coherence length~\cite{cerbino2008}. Over this distance, the distortion of the speckle interference pattern is only governed by the wavefront distortions. In the hard X-ray regime, this distance ranges from several millimeters to many meters depending on the source divergence and coherence, thereby allowing the small deviation angles coming from the wavefront distortions to be measured.

The speckle pattern used to modulate the wavefront can be obtained either through an interferometric process, meaning that the random pattern is a true speckle pattern, or from an absorption mask with random small apertures. In practice, the speckle observed with X-rays is often a combination of interference and absorption processes.

The experimental simplicity of the speckle-based techniques comes at a higher computational cost with respect to alternative techniques which are more instrumentation-demanding. Speckle-based techniques rely on tracking similar or resembling signals either in different propagation planes or at different times or configurations as it is the case for the differential metrology mode~\cite{berujon2012prl}. Since each speckle is or can be assumed to be unique, each grain acts as a marker allowing to determine the ray's propagation direction in a geometrical way. Using a series of detector images with a slightly shifted speckle generator or membrane increases the resolution provided by the tracking principle up to that of the detector and also offers an increased robustness.

Here, for the signal tracking, i.e. in order to follow or to find the size of the lateral displacement of a speckle on the detector plane, we will solely use the originally introduced method that searches for the maximum of a correlation peak~\cite{berujon2012prl, morgan2012}. Since then, other authors have employed different algorithms based on the minimization of a square sum difference~\cite{zanette2014, vittoria2015, zdora2017} with or without randomness in the wavefront modulator. Nevertheless, these numerical recipes are equivalent if we set the dark-field signal issue aside.

Let us assume two 'speckle' signals $(s_1,s_2)$, or random sets of two-dimensional intensity patterns on a detector, that are functions of $(\mathbf{r},\tau)$. The parameter $\mathbf{r}$ represents a pixel position on the detector in the $(x,y)$ basis transverse to the  X-ray beam propagation direction $z$. The controllable parameter $\tau$ describes the position in $(x,y)$ of the speckle generator, which can be moved perpendicularly to the beam direction, using high precision motors. Both signals $(s_1,s_2)$ can thus have a dimension up to four: two dimensions for the 2D position of a point on the detector $\mathbf{r}(x,y)$ and two other dimensions for $\tau(x,y)$. The signals are collected at different propagation distances, e.g. $s_1=s(\mathbf{r},\tau,z=z_1)$ and $s_1=s(\mathbf{r},\tau,z=z_2)$ in the absolute mode, or when the detector is kept fixed with $z_1=z_2$ while the beam state has been changed by, for instance, the insertion of an object in the differential mode.

Similarity in the signals is found numerically by calculating the normalized cross-correlation coefficients $C$ which can be written in the most general case:
\begin{equation}
C(\mathbf{v},\mathbf{u}) = \iint \widetilde{s_1}(\mathbf{r+v},\mathbf{\tau+u})\widetilde{s_2}(\mathbf{r},\mathbf{\tau}) d\mathbf{r}d\mathbf{\tau}
\label{eq:ZNCC}
\end{equation}

The operator $\widetilde{s}$ denotes a signal normalization over the window function $w_{r_0,\tau_0}(\mathbf{r},\mathbf{\tau}) = w(\mathbf{r -r_0})w(\mathbf{\tau - \tau_0)}$ centered around $\mathbf{r_0}$ and $\mathbf{\tau_0}$ and fulfilling the normalization condition:
\begin{equation}
\iint w_{r_0,\tau_0}(\mathbf{r},\mathbf{\tau})d\mathbf{r}d\mathbf{\tau} = \int w_{r_0}(\mathbf{r}) d\mathbf{r}= \int w_{\tau_0}(\mathbf{\tau})d\mathbf{\tau} =1
\end{equation}

Whilst $w_{r_0}(\mathbf{r})$ is used to define the position of the detector, i.e. where the calculation is performed, $w_{\tau_0}$ is usually defined by the range of speckle generator positions that are scanned. In practice, a window $w_{r_0}$ with larger dimensions will result in a lower spatial resolution as a trade-off for a greater accuracy gain. For $w_{\tau_0}$, a larger window corresponds to increased data collection but also to higher accuracy.

We have then for $s(\mathbf{r},\mathbf{\tau})$:
\begin{equation}
\widetilde{s}(\mathbf{r_0},\mathbf{\tau_0}) = \frac{w_{\mathbf{r_0,\tau_0}}s - \overline{s}}{\sigma(s)}
\end{equation}
with $\overline{s}$ the mean of $s$ over the window:
\begin{equation}
\overline{s}(\mathbf{r_0},\mathbf{\tau_0}) = \iint w_{\mathbf{r_0,\tau_0}} s~d\mathbf{r}d\mathbf{\tau}
\end{equation}
and $\sigma(s)$ its standard deviation:
\begin{equation}
\sigma[s(\mathbf{r_0},\mathbf{\tau_0})] = \sqrt{\iint \left( w_{\mathbf{r_0,\tau_0}} s - \overline{s} \right)^2 d\mathbf{r}d\mathbf{\tau}}
\end{equation}

As indicated above, some authors prefer to use an alternative operator instead of the cross-correlation one \cite{zanette2014,zdora2017}, called Normalized Sum of Square Difference, $B$, that must be minimized:
\begin{equation}
 B(\mathbf{v},\mathbf{u}) = \iint \left(\widetilde{s_1}(\mathbf{r+v},\mathbf{\tau+u}) - \widetilde{s_2}(\mathbf{r},\mathbf{\tau}) \right)^2 d\mathbf{r}d\mathbf{\tau}
\end{equation}

The relation between the operators $C$ and $B$ is simply:
\begin{equation}
B = 2\left[ 1 - C \right]
\end{equation}
if, again, one does not take into account the dark-field signal.

The different speckle approaches, explained in the following sections, were initially introduced within the extreme cases of $w_{\tau_0}(\mathbf{\tau}) = \delta (\tau-\tau_0)$ for XST or  $w_{r_0}(\mathbf{r}) = \delta (r-r_0)$ for XSS, where $\delta$ denotes the Dirac distribution. However, improved methods were later established by using larger window functions and hence providing an information gain from the neighboring pixels as well as integrating redundant information, which helped increase the methods' sensitivity and noise robustness.

The displacement vector $\upsilon$ from $(s_1,s_2)$ for a pixel at position $\mathbf{r_0}$ on the detector is defined as:
\begin{equation}
\upsilon=(\mathbf{v}_0+\mathbf{u}_0) = \argmax_{(\mathbf{v},\mathbf{u})} C = \argmin_{(\mathbf{v},\mathbf{u})} B
\label{eq:argmax}
\end{equation}

The tracking of the speckle signals is used to infer geometrically the trajectory of the X-rays and to recover the wavefront $W$. The relation is valid within the first Born approximation and in a homogeneous media of optical index $n$. In that case, $n\frac{d\mathbf{p}}{d\mathbf{z'}} = \textrm{grad}~W$, where $\mathbf{p}$ is a position vector of a typical point on the ray with direction $\mathbf{z'}$~\cite{born}. The beam phase $\phi$ is hence linked to the wavefront through the approximate geometrical view: $\phi = k W$, $k$ being the wavenumber.

\begin{figure}
\includegraphics[width=8cm]{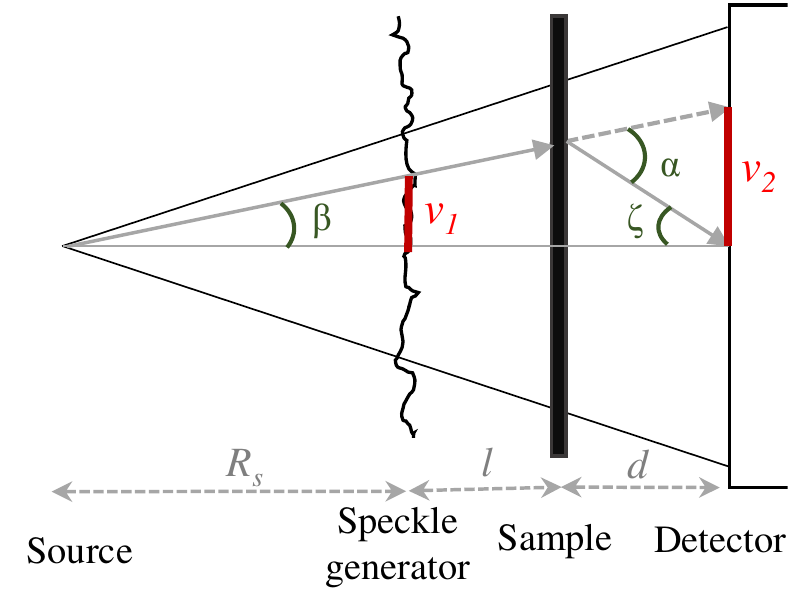}
\caption{Illustration of the origin of the $\Gamma$ factor for the case of a diverging source and a speckle generator located upstream of the sample. The photon deflection angle induced by refraction in the sample is represented by $\alpha$ (located in the sample plane). When the displacement vector corresponds to the vector illustrated by $\upsilon_1$ (in the case of XSS) rather than $\upsilon_2$ located on the detector (for instance with XST or XSVT), one must scale the displacement vector by a $\Gamma$ factor to recover $\alpha$ through the relationship linking the angles: $\alpha = \beta+\zeta$.
Equivalent considerations can be used when employing a setup where the speckle generator is placed after the sample.\label{fig:gamma}
}
\end{figure}

Using the notation $\nabla$ for the del (nabla) operator, the beam phase and wavefront can be linked to the displacement vector by:
\begin{equation}
\alpha= \nabla \phi= k\nabla W=\Gamma k\frac{\upsilon}{d}
\label{eq:pgrad}
\end{equation}
where $\Gamma$ is a geometric magnification scalar, cf. Fig. \ref{fig:gamma} and \citet{berujon2015pra}, and $d$ is the relevant propagation distance previously defined.

This signal tracking principle is implementable in different geometries and adaptable for different applications. The sample can be placed either upstream or downstream of the speckle generator whilst the detector position can be varied in several manners. Next, we briefly review the different techniques based on the signal tracking principle.

\subsubsection{X-ray speckle tracking (XST)\label{sec:XST}}

X-ray speckle tracking (XST) in differential and absolute modes was the first near-field speckle-based processing approach described in the literature~\cite{berujon2012prl, morgan2012}. It is directly inspired from numerical recipes employed in mechanics and particle image velocimetry where digital image correlation algorithms are intensively used to infer mechanical deformation or fluid fields \cite{tozzi2016,raffelbook}.

XST consists of recording only two images of the same speckle pattern with and without a sample inserted into the beam, for the differential mode, or at two propagation distances in the case of the absolute mode, the speckle generator being kept static at all times. The first image represents the signal $s_1(\mathbf{r})$ and the second one the reference signal $s_2(\mathbf{r})$. Given that this method is not a scanning technique, $s_1$ and $s_2$ are independent of $\tau$.
The displacement vector is simply recovered from Eq.~\ref{eq:ZNCC} and \ref{eq:argmax}:

\begin{equation}
\begin{aligned}
\upsilon&=\mathbf{v}_0 = \argmax_{\mathbf{v}} C_{XST}[s_1(\mathbf{r}),s_2(\mathbf{r})]\\
&=\argmax_{\mathbf{v}}\iint \widetilde{s_1}(\mathbf{r+v})\widetilde{s_2}(\mathbf{r}) d\mathbf{r}
\label{eq:xst1}
\end{aligned}
\end{equation}

where $\mathbf{u}$ is equal to zero and $w_{r_0}{\mathbf{r}}$ has a typical size of about 7 to 20 pixels in diameter. Here the spatial resolution of the technique is lower than that of the detector. However, the need of a single exposure of the sample makes it more readily applicable to time-resolved or dynamical studies.

In the differential mode, XST is quite straightforward to apply, with two images being taken with and without the sample inserted into the beam, the detector remaining in a fixed position.

In this case, $\Gamma $ from Eq.~\ref{eq:pgrad} is equal to 1 and $\nabla \phi = k\mathbf{v}_0/d$ in the plane of the sample when the speckle generator is placed upstream of the sample. For the case where the positions are inverted, the $\Gamma$ factor must be adjusted accordingly. Using the notation $R_s$ for the distance from the source to the sample and $l$ for the distance between the speckle generator and the sample, then $\Gamma = 1+l/R_s$.

XST in the absolute mode was first described in \citet{berujon2012prl}. In this mode, the two images are taken at different distances downstream of the speckle generator and subsets of the first image are tracked across the second one. The speckle grains are used as markers materializing the trajectory of the X-rays between the two propagation planes. The setup is depicted in Fig.~\ref{fig:geomabs} where the detector is located at two distinct positions separated by a distance ranging from a few tens of millimeter for photons of a few kilo-electron volts energy to a few meters when working at a higher energy, a typical distance being of a few hundreds of millimeters.

In the XST absolute mode, the subsets tracked from one image to the other image can be strongly distorted due to the setup geometry. Whilst the tracking of the correlation peak, as defined by Eq.~\ref{eq:xst1}, does not take into account the subset distortion induced by the sample, more robust methods exist, consisting of maximizing $C_{XST}$ of Eq.~\ref{eq:xst1} as a function of six or more variables. These algorithms take into account the subset distortions and improve the method robustness and accuracy although at greater cost in terms of computing resources~\cite{berujon2015josr}.

The XST principle was also demonstrated for single pulse metrology using an experimental setup in which a semi-transparent X-ray detector is followed by a second downstream detector~\cite{berujon2015josr,berujon2017spie} thus allowing images to be recorded on both detectors simultaneously. This experimental setup permits to track temporally the absolute variation of the beam.

Recently, a fast Fourier transform based algorithm was shown to be able to quickly recover the speckle displacement for a whole image with high resolution~\cite{paganin2018}. However this mathematical processing displays a singularity at the zero-frequency position in the Fourier space that requires a regularization process for the formula to be stable. Unlike for imaging where small features matter the most, and particularly for computed tomography where low spatial frequencies are usually filtered out in the reconstruction process, this regularization is problematic in the case of metrology applications since we are mainly interested in the low order aberrations. The use of real space methods avoids this amplification of low frequency noise.

\subsubsection{X-ray speckle vector tracking (XSVT)\label{sec:xsvt}}

The X-ray speckle vector tracking (XSVT) processing scheme was first described in the second part of \citet{berujon2017pra}. In this scanning technique, two sets of multiple images are collected at $N$ different transverse positions (initially randomly chosen) of the speckle generator $\tau(x_i,y_i), i\in [[1,N]]$. During the acquisition of the first set, the sample is inserted into the beam. This set of images constitutes the signal $s_1$ which is reorganized into a three dimensional array by stacking the images. Then, the signal $s_2$ is built repeating the very same scan, however this time with the object of investigation removed from the beam or at a different propagation distance $z$.

Although $s_1$ and $s_2$ are three-dimensional, the correlation is operated over $\mathbf{r}$ and with $\mathbf{u} = 0$:
\begin{equation}
\begin{aligned}
\upsilon&=\mathbf{v}_0 = \argmax_{\mathbf{v}} C_{XSVT}[s_1(\mathbf{r,\tau}),s_2(\mathbf{r,\tau})]\\
&=\argmax_{\mathbf{v}}\iint \widetilde{s_1}(\mathbf{r+v},\mathbf{\tau})\widetilde{s_2}(\mathbf{r},\mathbf{\tau}) d\mathbf{r}d\mathbf{\tau}
\label{eq:xsvt1}
\end{aligned}
\end{equation}
As for XST, $\Gamma =1$ if the membrane is located upstream of the sample, or $\Gamma = 1+l/R_s$ if the membrane is downstream.
When the method was first introduced, XSVT used $w_{\mathbf{r_0}}(\mathbf{r -r_0}) = \delta(\mathbf{r -r_0})$, $\delta$ being the Dirac delta function. Later, the technique evolved with the use of a small $w_{\mathbf{r_0}}$ window, losing some of the resolution as a trade-off for getting more accuracy without the need to increase the amount of data collected~\cite{berujon2016}. In order to avoid computing time issues, one should select a window size such that the total number of elements in the vector does not exceed 100. Whilst for metrology purposes a large number of images $N$ and a unitary $w_{\mathbf{r_0}}$ window size are preferable, for imaging applications one aims at using only a few images while increasing the $w_{\mathbf{r_0}}$ window size until sufficient statistics can be obtained, e.g resulting in a typical size for $w_{\mathbf{r_0}}$ of 3x3 to 6x6.
XST can be seen as a special case of XSVT with $w_{\mathbf{\tau_0}}(\mathbf{\tau -\tau_0}) = \delta(\mathbf{\tau -\tau_0})$.

The considerations made for XST in the absolute and differential metrology modes can be applied to XSVT in an equivalent manner. By recording the signals $(s_1, s_2)$ at different propagation planes, one can access the absolute metrology in contrast to the differential metrology obtained when signals are recorded at the same location but at different times or in different conditions.

\subsubsection{X-ray speckle scanning (XSS)\label{sec:XSS}}

The X-ray speckle scanning (XSS) approach was first described in \citet{berujon2012pra} and later simplified in terms of data collection in \citet{berujon2017pra}. It was shown that this processing method is equivalent to the one used in grating interferometry where one of the gratings is scanned in the phase stepping mode.
In this case, the signals $(s_1,s_2)$ are built by scanning the speckle generator position $\tau$ using a fine mesh.
The signals $(s_1,s_2)$ are four-dimensional and the cross-correlation is done over $\tau$ with $\mathbf{v} = 0$:
\begin{equation}
\begin{aligned}
\upsilon&=\mathbf{u}_0 = \argmax_{\mathbf{u}} C_{XSS}[s_1(\mathbf{r,\tau}),s_2(\mathbf{r,\tau})]\\
&=\argmax_{\mathbf{u}}\iint \widetilde{s_1}(\mathbf{r},\mathbf{\tau+u})\widetilde{s_2}(\mathbf{r},\mathbf{\tau}) d\mathbf{r}d\mathbf{\tau}
\label{eq:xss1}
\end{aligned}
\end{equation}
For XSS, the $\Gamma $ factor of Eq.~\ref{eq:pgrad} depends on the geometry and the divergence of the setup. As one can see on Fig. \ref{fig:gamma}, the displacement vector calculated with XSS corresponds to $\upsilon_1$, a length located in the plan of the speckle generator. The $\Gamma$ factor permits here to scale the angle $\alpha$ as a function of $\upsilon_1$. With a source located at a distance $R_s$ and the object placed at a distance $l$ from the speckle generator, it was demonstrated that $\Gamma =  (R_s+l+d)/R_s$ for scaling the angles at the sample position \cite{berujon2015pra}. When a collimated probe beam is used, $\Gamma$ is reduced to 1.

Another XSS scheme often used is based on a one dimensional scan followed by the use of a highly asymmetric window $w(\mathbf{r-r_0})$ for the cross-correlation calculation. This variant, often presented as XSS-1D, is equivalent to XSS in the dimension orthogonal to the scan direction and to the XSVT method in the other beam transverse direction \cite{berujon2012pra,wang2016}. For a scan in the $\mathbf{x}$ direction, the previous equation is transformed for this method into:
\begin{equation}
\begin{aligned}
\mathbf{u_x}_0,\mathbf{v_y}_0 = \argmax_{\mathbf{u,v}}\iint \widetilde{s_1}(\mathbf{r+\mathbf{v_y}},\mathbf{\tau+u_x})\widetilde{s_2}(\mathbf{r},\mathbf{\tau}) d\mathbf{r}d\mathbf{\tau}
\label{eq:xss1d}
\end{aligned}
\end{equation}
The displacement vector is hence separated into orthogonal components by forcing the use of the projections $\mathbf{u_x}=\mathbf{u} \cdot \mathbf{x}$ and $\mathbf{v_y}=\mathbf{v}\cdot \mathbf{y}$.

In \citet{berujon2017pra}, a variant of the XSS processing scheme showed ways of reducing the amount of data collected in an attempt to infer each time injective functions of $s_1$ to $s_2$. The strategy explained therein mainly consists of collecting a complete reference scan by performing fine, though limited in size, mesh scans of $s_2$ while recording a lesser number of points for $s_1$ with the speckle generator positioned at the central points $P_k$ of the previous meshes. In that case, the reorganized data for $s_1(\mathbf{r},\mathbf{\tau},P_k)$ and $s_2(\mathbf{r},\mathbf{\tau},P_k)$ go up to dimension five, the correlation simply becoming:

\begin{equation}
C_{XSS}(s_1,s_2) =\frac{1}{N}\sum_k^N\iint \widetilde{s_1}(\mathbf{r},\mathbf{\tau+u},P_k)\widetilde{s_2}(\mathbf{r},\mathbf{\tau},P_k) d\mathbf{r}d\mathbf{\tau}
\label{eq:xss2}
\end{equation}

When XSS was first presented in \citet{berujon2012pra}, the processing was done pixel by pixel, i.e. with $w_{\mathbf{r_0}}(\mathbf{r -r_0}) = \delta(\mathbf{r -r_0})$. However, as in the case of XSVT, a gain in accuracy could be achieved by enlarging the window $w_{\mathbf{r_0}}$ at the cost of a reduced spatial resolution \cite{berujon2017pra}. Similarly to XSVT, a simple way of numerically implementing this feature consists of concatenating data from different pixels of $s$ along a fourth or fifth dimension. However, this numerical processing, although trivial to program, can quickly become memory intensive if no care is taken.


\subsubsection{Speckle scanning self-correlation mode\label{sec:XSSself}}

In the self-correlation speckle scanning processing method, a variant of XSS, only one large 1D or 2D scan of the speckle generator is performed, with the optics in the beam. Although this method gives access to the wavefront curvature that includes the defects of the probe beam, here the distance $d$ is that from the sample to the detector as shown in Fig.~\ref{fig:geomdiff}. In this method, $s_1 = s_2$ and a self-correlation of this signal among different pixels is operated:
\begin{equation}
\begin{aligned}
\upsilon&=\mathbf{u}_0 = \argmax_{\mathbf{u}} C_{XSSabs}[s_1(\mathbf{r,\tau}),s_1(\mathbf{r + \Delta r,\tau})]\\
&=\argmax_{\mathbf{u}}\iint \widetilde{s_1}(\mathbf{r}+ \Delta r,\mathbf{\tau+u})\widetilde{s_1}(\mathbf{r},\mathbf{\tau}) d\mathbf{r}d\mathbf{\tau}
\label{eq:xssself}
\end{aligned}
\end{equation}
From the calculation of $\upsilon$, one can deduce the local absolute beam curvature $\kappa = R^{-1} = \partial^2 W/\partial \mathbf{r^2}$ (in the detector plane) using the relation of the magnification:
\begin{equation}
\Gamma = \frac{\kappa^{-1}}{\kappa^{-1}-l-d}=\frac{\mathbf{\Delta r}}{\upsilon}
\end{equation}
Which gives:
\begin{equation}
\kappa = \frac{1 - \frac{\upsilon}{\mathbf{\Delta r}}}{l+d}\label{eq:abscurv}
\end{equation}
Minimizing $l$ of the experimental setup or using a collimated probe beam, the expression reduces usually to $\kappa = \frac{1 - \frac{\upsilon}{\mathbf{\Delta r}}}{d}$.

In practice, $\Delta r$ is on the order of a few pixels and $w_{r_0}$ is often equal to $w_{r_0}(\mathbf{r}) = \delta (r-r_0)$. This processing scheme assumes a well-known and constant pixel size over the full detector active area. This parameter can be extracted using speckle-based methods as explained in \citet{berujon2015josr}.
Hence, by calculating four curvature fields, the absolute wavefront maps can be recovered through a double integration.

Within the small angle approximation, the curvature fields being linear, the differential phase induced by a sample can also be recovered by integrating a difference of curvature. One can obtain the curvature fields of two optical configurations by performing two XSS self-correlation scans and the difference can be obtained by simple subtraction if the beam did not move too much between scans.

As with the XSS technique, one can proceed by operating a one dimensional scan and use an asymmetric window in the correlation operation. That will again allow to reduce the data acquisition time at the cost of a moderate loss in spatial resolution in the dimension orthogonal to the scan direction.
This XSS self-correlation 1D scheme is highly beneficial in order to reduce the data collection time when the problem can be projected in the orthogonal directions (e.g. for the focusing of a KB system).

As we shall show later, this processing method is found to be of interest for measuring optics that are strongly distorting the wavefront.

\subsubsection{Hybrid tracking: XSVT-XSS\label{sec:XSVT-XSS}}

While the XSVT and XSS methods restrict the search of $v$ by cross-correlations solely over either the $\mathbf{r}$ or $\mathbf{\tau}$ parameters, we can come back to Eq.~\ref{eq:argmax} in which the cross-correlation is operated over both $\mathbf{v}$ and $\mathbf{u}$. Now, let us consider the signals $s_1$ and $s_2$ collected during scans of the speckle generator whose range is much larger than a demagnified pixel size $p_{ix}/\Gamma$.

We now consider $\mathbf{v} \neq \mathbf{u} \neq 0$ and $v = \mathbf{v} + \mathbf{u}$, as in the general case. In practice, as for XSS, the hybrid method merging XSS and XSVT starts with the recording of $s_1$ and $s_2$ by operating two bidimensional scans of the speckle generator. The signals from different pixels are numerically correlated to track their shift over the two variable parameters $\mathbf{v}$ and $\mathbf{u}$ using:
\begin{equation}
\begin{aligned}
v&=\mathbf{u}_0 +\mathbf{v}_0= \argmax_{\mathbf{u},\mathbf{v}} C_{XSShyb}[s_1(\mathbf{r,\tau}),s_2(\mathbf{r,\tau})]\\
&=\argmax_{\mathbf{v},\mathbf{u}}\iint \widetilde{s_1}(\mathbf{r+v},\mathbf{\tau+u})\widetilde{s_2}(\mathbf{r},\mathbf{\tau}) d\mathbf{\tau}d\mathbf{r}
\label{eq:xsshyb}
\end{aligned}
\end{equation}

Although slightly more complicated to implement and more demanding in terms of computation time, this processing method can be useful when high sensitivity is required for strongly curved wavefronts, both in absolute and differential modes. When used within a magnifying setup, the $\Gamma$ factor of the method has two distinct components for $\mathbf{v}$ and $\mathbf{u}$.

\begin{table*}
\begin{center}
\includegraphics[width=1.2\textwidth, angle=90]{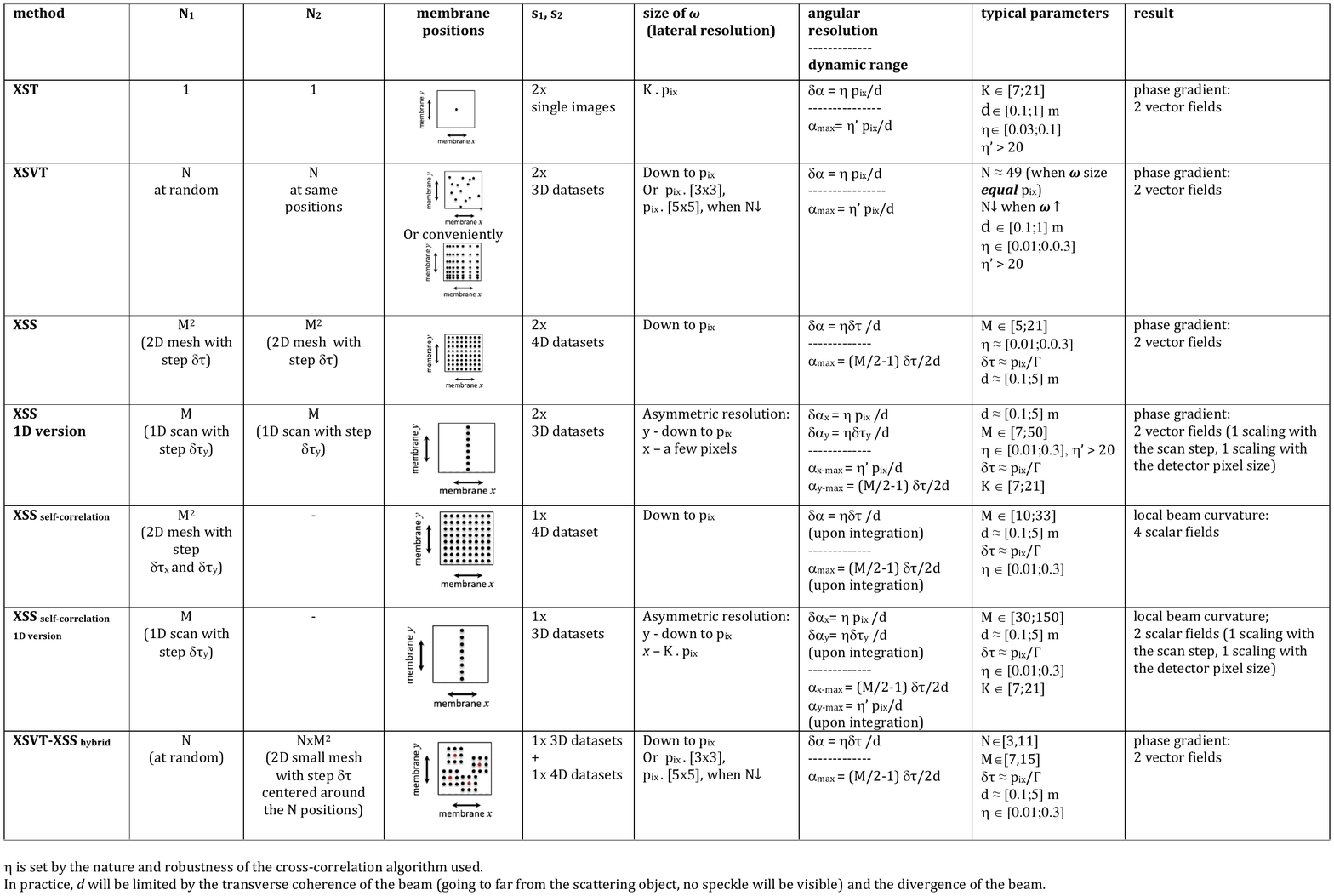}\\
\end{center}
\textbf{Table 1\label{tab:comp}}\\
\footnotesize{Comparative table of the different techniques.}
\end{table*}

Note that, as with the other processing methods, the size of $w_{r_0,\tau_0}$ can be larger than unity for both parameters, thus increasing the amount of data injected in the correlation calculation, and requiring further computing resources.

\section{Relevance of the methods}

Inferring from theoretical considerations, one can evaluate the differences between the various processing schemes, thus assessing where the use of a particular scheme becomes more relevant with respect to the others.

When phase sensing is used for imaging, the context asks for the lowest number of exposures per projection. In imaging, one is often willing to sacrifice some of the quantitative accuracy for a gain in visual quality. In a computed tomography scan, hundreds to thousands of projection images are collected around a sample to numerically reconstruct its volume. Algorithms with a single exposure per angular projection are thus highly desirable  when working with sensitive materials in order to keep both the scanning times and the deposited dose as low as possible. The algorithm proposed by Paganin \emph{et al.} \cite{paganin2018} is for instance very well suited for this purpose. Additionally, the ramp filter used in the tomographic reconstruction combines with the one for the 2D numerical integration in Fourier space to form a phasor to eventually greatly reduce the artifacts generated independently by each filter. As mentioned earlier in this manuscript, the previously mentioned algorithm is not as well suited to metrology applications as for imaging. Operating in Fourier space requires dealing with the singularity of the filter around the zero frequency component. Any regularisation of these low frequencies in a projection mode will likely generate low frequency artefacts which are dramatic for metrology applications.

In contrast to imaging purposes, metrology applications call for a higher sensitivity and quantitative accuracy even if this leads to longer scans. For most optical characterization, this however does not pose a problem.

Table 1 offers a quick overview of the schemes' parameters and characteristics.

The XST technique offers the poorest resolution among the presented approaches since it uses a whole subset of pixels to map the trajectory of a ray. However, given that the technique only requires two images (of which one acts as a reference), it permits to monitor dynamic processes. As we shall see, the technique can prove useful for stability assessments. On the downside, XST can sometimes suffer from numerical artefacts generated by absorbing or diffracting elements in the beam or by a wavefront highly perturbed near the spatial frequencies of the speckle grains.

XSVT is an attractive technique that provides both a good resolution (which can be as good as that of the detector) and a good performance, with robustness to noise due to the redundancy of the data. More importantly, XSVT is a very versatile technique that provides an almost infinite dynamic range in angular sensitivity while offering an opportunity to tune the scan length to obtain the desired accuracy and resolution. 

The XSS technique is experimentally the slowest one, since it requires long scans ($N^2$ images when applied in two dimensions). On the other hand, but at the same time it provides the best angular sensitivity both in the differential and self-correlation modes. This is due to the smallest displacement vector $\mathbf{v}_\textrm{min}$ that can be sensed with this technique, equal to a fraction of the demagnified detector pixel ${\mathbf{v}}_\textrm{min}<p_{ix}/\,\Gamma$. Since XSS allows $\Gamma$ to be very large, smaller displacement vectors and angles can be sensed down to a single nanoradian.

The XSS self-correlation mode is also of interest since it can provide the absolute local curvature of the beam with high angular precision and without the need of moving the detector during the measurements. As we shall demonstrate in an accompanying paper, this technique offers high performance for characterising the absolute wavefront and provides a robust alternative when differential measurements are made impossible due to strongly focusing optics.


The hybrid technique allows one to combine the strength of both the XSVT and XSS approaches, yet at the cost of a higher complexity in the numerical implementation.

In a following complementary manuscript, we will present several experimental applications and discuss the choice of techniques depending on each metrology scenario.

\section{Conclusions} 

Several aspects of speckle tracking using various geometries and processing schemes have been reviewed. X-ray near-field speckle methods can be used to extract both absolute and differential characteristics of the X-ray beam wavefront, thus allowing to characterise optics using at-wavelength metrology. The configurations presented here are not exhaustive but they permit to illustrate the core principle of the methods and the way they relate to each other. The combination of a setup and a processing scheme can be adapted to operate optimally for characterising a wide range of X-ray optical elements.

In a subsequent paper, experimental applications will be detailed in order to illustrate the potential of the reviewed methods for a variety of situations and purposes. For instance, we will demonstrate how the methods can provide important information regarding the performance and the defects of refractive and reflective optics. While the presented data were obtained at a synchrotron, it is worth mentioning that the techniques are expected to work with other kinds of X-ray sources~\cite{zanette2014}, at a variety of energies~\cite{wang2016} and even with visible light~ \cite{berto2017}.

\vspace{0.5cm}
The authors wish to thank the ESRF for financial and personal support.


%

\end{document}